\documentstyle[pre,multicol,aps]{revtex}
\input{epsf.tex}

\begin{document}
\title{Intensity limits for stationary and interacting multi-soliton complexes}

\author{Andrey A. Sukhorukov}
\address{Nonlinear Physics Group, 
Research School of Physical Sciences and Engineering,\\
Australian National University, Canberra, ACT 0200, Australia}

\author{Nail N. Akhmediev}
\address{Optical Science Centre,
Research School of Physical Sciences and Engineering,\\
Australian National University, Canberra, ACT 0200, Australia}

\maketitle
\begin{abstract}
We obtain an accurate estimate for the peak intensities of multi-soliton
complexes for a Kerr-type nonlinearity in the (1+1) -- dimension problem. Using exact analytical solutions of the integrable set of nonlinear Schr\"odinger equations, we establish a rigorous relationship between the eigenvalues of incoherently--coupled fundamental solitons and the range of admissible intensities. A clear geometrical interpretation of this effect is given.
\end{abstract}


\vspace*{-1cm}
\begin{multicols}{2}
\narrowtext

Pulse or beam evolution under the action of competing linear dispersion and nonlinear self-action is a fundamental phenomenon which can be observed in optical fibers, nonlinear crystals or gases, atomic Bose-Einstein condensates, etc. The corresponding wave properties can often be described by {\em the nonlinear Schr\"odinger equation} (NLSE). In the (1+1)-dimensional case, if the homogeneous medium has a Kerr-type nonlinear response, this equation is integrable~\cite{ZS}, and its solutions can be found analytically. They describe a nonlinear superposition of localized waves, called {\em solitons}, and radiation modes. Such analysis allows us to reveal some general features of nonlinear localized waves.

Localized solutions of the NLSE are among the main objects of interest for telecommunication and information processing applications. All possible localized solutions of integrable systems are composed of solitons. Hence it is important to understand their interactions and collective properties. For a single NLSE describing the evolution of a coherent field, soliton interactions are phase-sensitive, so that constructive or destructive nonlinear interference can be observed. Since the profiles of interacting solitons are distorted according to the nonlinear superposition principle, large variations of the peak intensities can occur. Indeed, it has been demonstrated that the peak intensity can vary by a factor of $N^2$ for $N$ interacting solitons~\cite{AA38}. In some sense, phase sensitivity can be ``amplified'' due to nonlinearity, so that it is greater than in the case of linear interference between mutually coherent sources.

Interactions of fundamental solitons can also be {\em phase-insensitive} if they correspond to {\em incoherently coupled} components. Then, fundamental solitons can be composed into a stationary multi-soliton complex (MSC). To have a rough idea of an MSC, we can use the following analogy. If we think of solitons as elementary particles, then we can imagine an MSC as a more complicated structure, such as an atom. An optical incoherent soliton is an example of an MSC. These objects have been studied extensively both experimentally and theoretically during recent years (see Refs.~\cite{Incohfirst,Vysl2,mprl,IncohApproaches} and references therein); some other examples of MSCs are listed in Ref.~\cite{review}. The question of the maximal intensity of stationary and interacting MSCs is of primary interest for applications. However, to the best of our knowledge, this problem has not been addressed yet. In this work, we derive a rigorous relationship between the parameters of incoherently--coupled solitons and the range of admissible intensities for MSCs.

The evolution of spatial incoherent solitons in a medium with a Kerr-type self-focusing nonlinearity can be described by a set of NLSEs for mutually--incoherent components~\cite{IncohApproaches}:
\begin{equation} \label{eq:nlse}
  i \frac{\partial u_{m}}{\partial z}
  + \frac{1}{2} \frac{\partial^{2}u_{m}}{\partial x^{2}}
  + I u_m
  = 0 ,
\end{equation}
where $u_m$ is the normalized amplitude of the $m$-th component ($1 \le m \le M$), $M$ is the number of components, $I = \sum_{j=1}^M |u_j|^2$ is the total intensity, $z$ is the coordinate along the direction of propagation, and $x$ is the transverse coordinate (we are considering a (1+1)-dimensional case).

Eqs.~(\ref{eq:nlse}) are {\em completely integrable} by means of the inverse scattering technique (IST)~\cite{Manakov} and, therefore, any localized solution can be represented as a nonlinear superposition of solitary waves. Every {\em fundamental soliton} (labeled~$j$) is characterized by a complex eigenvalue $k_j = r_j + i \mu_j$, and a shift in the coordinate plane $(x_j,z_j)$. The simplest single--soliton solution can be written as: $u_j(x,z) = r_j \; {\rm sech}( \beta_j ) \; e^{i \gamma_j}$, where $\beta_j = r_j (\bar{x}_j - \mu_j \bar{z}_j)$, $\gamma_j = \mu_j \bar{x}_j + (r_j^2-\mu_j^2) \bar{z}_j / 2$, and $(\bar{x}_j,\bar{z}_j) = (x - x_j,z - z_j)$ are the shifted coordinates.
The solution for bright MSCs consisting of $M$ incoherently--interacting fundamental solitons can be found by solving a set of linear equations~\cite{Nogami,Gardner}:
\begin{equation} \label{eq:slau}
\sum_{m=1}^{M} \frac{e_j e_m^{\ast} u_m}{k_j +k_m^{\ast}}
+ \frac{1}{2 r_j} u_j = - e_j ,
\end{equation}
where $e_j = \chi_j \exp \left( \beta_j + i \gamma_j \right)$ and the coefficients $\chi_j$ define the relative coordinate centers for the fundamental solitons. The seemingly straightforward process of inverting the set of linear equations results in a complicated expression. 

Using the technique described in Ref.~\cite{mprl}, the solution of Eqs.~(\ref{eq:slau}) can be obtained in a compact explicit form. However, it is not possible to find a general analytical expression for the maximum intensity, and it can only be determined by numerically solving transcendental equations. Therefore, a different approach is needed to make an analytical estimate for the peak intensity levels. In order to do this, we first multiply Eqs.~(\ref{eq:slau}) by $u_j^{\ast} / (k_j - i \mu)$, add the complex conjugate, and sum over the fundamental soliton numbers $1 \le j \le M$ (we note that such a transformation was suggested earlier~\cite{mbackgr} as an intermediate step for deriving bright and dark soliton solutions on a background). Finally, we obtain the following inequality
\begin{equation} \label{eq:ineq}
  \sum_{j=1}^{M} \left| \frac{u_j}{k_j^{\ast} + i \mu} \right|^2
  \le 1 ,
\end{equation}
which is valid for any (real) $\mu$. This key result makes it possible to estimate the limitations on the total intensity. In particular, we conclude that the peak intensity is limited from above, i.e. $I_{max}(z) \equiv {\rm max}_x I(x,z) \le I_{inf}$, where the upper boundary is
\begin{equation} \label{eq:upper}
   I_{inf} = \max_j [r_j^2 + (\mu_j-\mu)^2] , 
\end{equation}
and $\mu$ is chosen to minimize the value of $I_{inf}$. We will now consider several examples illustrating the physical meaning of Eqs.~(\ref{eq:ineq}) and~(\ref{eq:upper}).

\begin{figure}[H]
\setlength{\epsfxsize}{8cm}
\centerline{\mbox{\epsffile{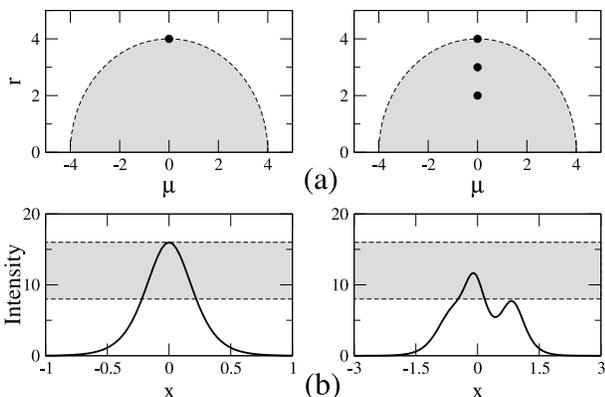}}}
\caption{ \label{fig:static}
(a)~Geometrical illustration of the maximum intensity criterion for a stationary MSC: black dots mark the eigenvalues of fundamental solitons in the $(r,\mu)$ parameter space, and the radius of the shaded half-circle determines the maximum amplitude.
(b)~The intensity profiles corresponding to upper plots; shading shows allowed ranges of peak intensities.
}
\end{figure}

{\em Stationary MSC.} 
Let us first analyse the properties of a single MSC, composed of several solitons with identical velocities, $\mu_n$. In such a case, the optimal choice of the free parameter in Eq.~(\ref{eq:upper}) is $\mu = \mu_n$, and we have $I_{inf} = \max_j (r_j^2)$. Note that this value can be interpreted as the minimal squared radius of a circle which has its center at the point $(0,\mu)$ in the parameter space $(r,\mu)$ and which contains all the soliton eigenvalues within it [see Fig.~\ref{fig:static}(a)]. Thus $I_{inf}$ is proportional to the area ($\pi R^2 / 2$) of this semi-circle. On the other hand, we note that for a stationary MSC the amplitude profiles satisfy a self-consistent eigenvalue problem, $d^2 U_m / d x^2 - r_m^2 U_m + 2 I\, U_m = 0$, where the functions $U_m = u_m \exp(-i \gamma_m)$ are real. Then, since solutions should be localized in the transverse direction ($x$), we conclude that $I_{max} > I_{sup} = \max_j (r_j^2 / 2)$. This means that the variations of the peak intensity are strictly limited by the largest eigenvalue in an MSC, as illustrated in  Fig.~\ref{fig:static}. Note that $I_{max} = I_{inf}$ if there is only one fundamental soliton (Fig.~\ref{fig:static}, left). The peak intensity decreases if several solitons compose an MSC, but always remains above the lower limit (Fig.~\ref{fig:static}, right). This happens despite the fact that the individual fundamental soliton intensities are always superimposed, i.e. destructive interference is not possible, in contrast to the case of coherent interactions. 
The observed {\em decrease of total intensity} underlines the complicated nature of the {\em nonlinear superposition} phenomenon and occurs because the profiles of individual solitons are strongly distorted due to the nonlinear self-action effect.

\begin{figure}[H]
\setlength{\epsfxsize}{8cm}
\centerline{\mbox{\epsffile{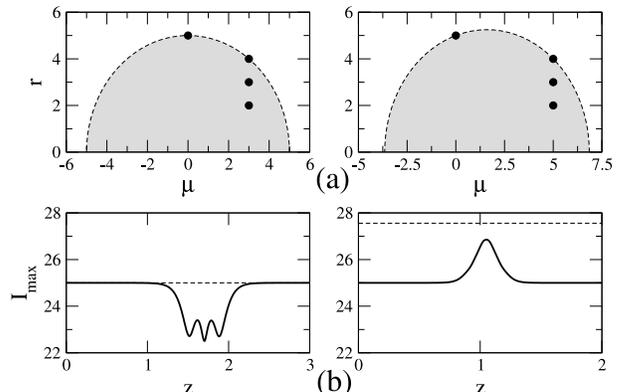}}}
\caption{ \label{fig:collis}
(a)~Geometrical illustration of the maximum intensity criterion for interacting MSCs: black dots mark the eigenvalues of fundamental solitons in the $(r,\mu)$ parameter space, and radius of the shaded half-circle determines the maximum amplitude.
(b)~Dependencies of the peak intensities, corresponding to the upper plots, on the propagation distance; dashed line shows the maximum possible value.
}
\end{figure}

{\em Interacting MSCs.} 
Our general results can also be applied to the case when solitons have different velocities, $\mu_n$. In other words, we can estimate the peak intensity changes during the collision of several MSCs. As follows from the form of Eq.~(\ref{eq:upper}), the limiting value $I_{inf}$ depends only on the {\em maximum eigenvalue in each of the colliding MSCs}. Again, Eq.~(\ref{eq:upper}) has a clear geometrical interpretation: the optimal value of $\mu$ in Eq.~(\ref{eq:upper}) must be chosen to minimize the area of a half-circle which has its center at the point $(0,\mu)$ and which contains all the soliton eigenvalues. Two examples, corresponding to a collision between a single fundamental soliton and an MSC with different velocities ($\mu_n$), are shown in Fig.~\ref{fig:collis}. When the relative velocity is small (Fig.~\ref{fig:collis}, left), the intensity at the impact area of the collision decreases. We have already observed this effect for a stationary MSC when the relative velocity is zero. However, for larger relative velocities, the peak intensity increases (Fig.~\ref{fig:collis}, right). 

\begin{figure}[H]
\setlength{\epsfxsize}{8.5cm}
\centerline{\mbox{\epsffile{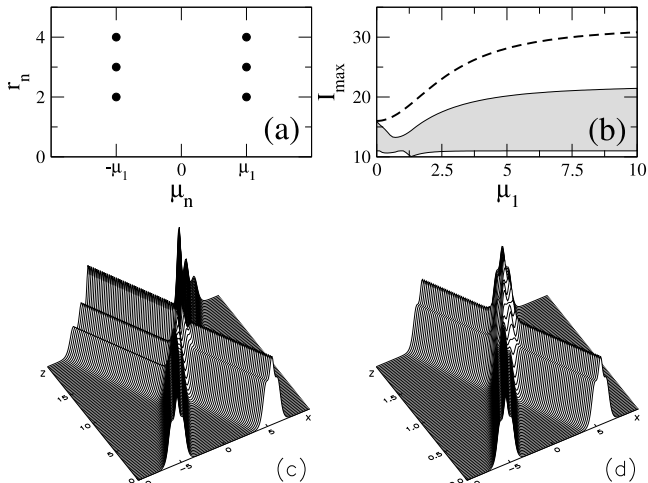}}}
\caption{ \label{fig:angular}
(a)~Location of the eigenvalues of fundamental solitons in the $(r,\mu)$ parameter space (black dots).
(b)~Ranges of the peak intensity variations during the soliton collision vs. relative velocity shown with shading; dashed line gives the maximum possible value.
(c,d)~Intensity profiles illustrating soliton collisions at small ($\mu_1=0.5$) and large ($\mu_1=5$) relative velocities.
}
\end{figure}

In order to understand the differences in the interaction pattern, we study the dependence of the peak intensities on the relative velocities of the colliding MSCs. To illustrate the key features, we consider interactions of two identical MSCs with a relative velocity $2 \mu_1$ [see Fig.~\ref{fig:angular}(a)]. Although Eq.~(\ref{eq:upper}) can be used to obtain an estimate for the peak intensities, as in the previous examples, we find that the results are not optimal for large $\mu_1$. To obtain a more accurate estimate, we recall that Eq.~(\ref{eq:ineq}) is satisfied for all (real) $\mu$ simultaneously. We now choose $\mu = \pm \mu_1$, add up the corresponding inequalities together, and obtain the following upper limit, $I_{inf} = \max_j [r_j^2 (r_j^2+4 \mu_1^2) / (r_j^2+2 \mu_1^2)]$. We thus see that $I_{inf}(\mu_1\rightarrow+\infty) \rightarrow 2 \max_j (r_j^2)$, which is a simple sum of the upper bounds for individual MSCs. This result means that interaction of solitons with large relative velocities is weak, and their intensities are added together similarly to the linear case. This conclusion is confirmed by numerical simulations, the results of which are presented in Fig.~\ref{fig:angular}(b). We also illustrate the evolution of the intensity profiles at small and large relative velocities in Figs.~\ref{fig:angular}(c) and~(d), respectively

Finally, we note that these results can be applied to bright solitons existing on top of a background~\cite{incoh_backgr_sf}, and also to dark solitons which can exist in media with a self-defocusing Kerr-type nonlinearity~\cite{incoh_backgr_df}. Indeed, it has recently been demonstrated~\cite{mbackgr} that the total intensity profile is uniquely determined by the eigenvalues of the fundamental solitons, and does not depend on the angular distribution of radiation modes which form the background. Therefore, in these cases, our estimates are valid for the maximum changes of intensity relative to the background level $I_b$, i.e. 
$\max_x |I(x,z) - I_b| \le I_{inf}$.

In conclusion, we have obtained an accurate estimate for the peak intensities of multi-soliton complexes for a homogeneous medium with a Kerr-type nonlinearity in the (1+1)--dimensional situation. We have demonstrated that incoherent coupling can result in a {\em decrease of peak intensities} when the relative soliton velocities are small (or zero) and nonlinear interaction is strong. On the other hand, solitons with large relative velocities, where the interaction is weak, can roughly be superimposed as linear waves.

The authors are members of the Australian Photonics Co-operative Research Centre (APCRC).
We are grateful to Adrian Ankiewicz for a critical reading of this manuscript.

\end{multicols}
\end{document}